\documentclass[aps,twocolumn, pra,showpacs]{revtex4-1}

\usepackage{epsfig,amssymb,amsmath,graphicx,color}
\usepackage[dvips]{hyperref}

\begin{document}

\title{Localization and entanglement through scattering measurements}
\author{James S. Douglas and Keith Burnett}
\affiliation{University of Sheffield, Western Bank, Sheffield S10 2TN, United Kingdom}
\begin{abstract}
We study the emergence of localization and entanglement in many-body systems as a result of scattering measurements. We show that consecutive scattering measurements on a many-body system can produce superposition states in position space. The resulting states are signaled in the scattering distribution and these states are robust under continued scattering. We also show that the average scattering distribution for an ensemble of experiments gives the initial state scattering distribution even when the initial state is completely destroyed by the scattering process.
\end{abstract}
%\date{29 July 2011}
\maketitle

Scattering of particles from matter is a fundamental method used to reveal a system's properties. A prominent example is Rutherford's revolutionary discovery of the atomic nucleus by scattering alpha particles from gold foil \cite{Rutherford1911a}. Scattering of neutrons has been used to probe the nature of exotic forms of matter such as superfluid helium and high temperature superconductors \cite{Griffin1993a}, while scattering of x-ray photons has led to numerous discoveries about crystalline structure, such as the structure of DNA \cite{Franklin1953a}, and scattering of optical photons has for example been used to probe correlations within Bose-Einstein condensates \cite{Stamper-Kurn1999a}.
Neutral probe particles scatter from matter either as a result of density fluctuations in the system or by creating excitation in the system. Measuring the scattering pattern then yields valuable information about the system's spatial structure and excitations, and furthermore interference between different scattering channels reveals information about correlations within the system.

Typically in experiments and theoretical discussions involving scattering the effect of the probe on the system is assumed to be weak and the scattering pattern is calculated in the Born approximation from the initial system state \cite{VanHove1954}. However during each scattering event energy and momentum may be transferred from the probe to the system. Furthermore the act of measuring the probe's scattering pattern may produce quantum backaction on the state of the system. These effects can combine to significantly alter the state of the system, indeed for two particles, scattering of light leads to localization of relative position while the particles' momentum states become entangled \cite{Rau2003a,Cable2005a}. 

Here we model the evolution of many-body states caused by consecutive scattering and measurement events, and show that for a system of multiple particles this evolution can lead to many-body superpositions in position space. Such states may be useful for making measurements with enhanced precision and for investigating the boundary between the classical and the quantum. Our model also shows that the average scattering pattern for an ensemble of experiments will revert to the pattern predicted for the initial state even if this initial state is completely destroyed in each individual scattering experiment. Experimental access to the initial state of system is hence possible even for systems that are easily perturbed. 

We begin by examining the scattering problem from a general perspective. We consider an atomic system of identical particles  with associated field creation and annihilation operators $\hat{\Psi}^\dagger(\mathbf{r})$ and $\hat{\Psi}(\mathbf{r})$. The probe particles are assumed to have a definite momentum with associated wave-vector $\mathbf{k}_0$. Scattering of the probe by the system is then associated with a momentum transfer to the system of $\hbar\mathbf{k} =\hbar(\mathbf{k}_0 - \mathbf{k}_f)$, where $\mathbf{k}_f$ is the probe wave-vector after scattering.
% and we assume $|\mathbf{k}_f|=|\mathbf{k}_0|$. 
Measurement of the final probe wave-vector, for example by imaging in the far field, then transforms the system state according to
\begin{equation}
|\psi\rangle \longrightarrow \int d\mathbf{r} \hat{\Psi}^\dagger(\mathbf{r})\hat{\Psi}(\mathbf{r})e^{i\mathbf{k}\cdot \mathbf{r}}|\psi\rangle.
\label{eq:scatter_transform}
\end{equation}
For an initial $N$-particle state $|\psi\rangle = \int d\mathbf{R} \phi(\mathbf{R})|\mathbf{R}\rangle$, where  $\mathbf{R}= \{\mathbf{r}_1,\ldots,\mathbf{r}_N\}$ gives the coordinates of the $N$ particles, the associated probability of observing a scattered probe wave-vector of $\mathbf{k}_f$ is
\begin{equation}
P(\mathbf{k}_f) = \frac{g^2}{4 \pi}\int d\mathbf{R}\left|\phi(\mathbf{R})\sum_{j=1}^N e^{i \mathbf{k}\cdot\mathbf{r}_j}\right|^2.
\label{eq:prob_scatter}
\end{equation}
The factor $g$ is determined by the strength of the probe-system interaction and is assumed to be small so that multiple scattering of an individual probe particle is negligible. 
For simplicity we assume $g$ is independent of $\mathbf{k}_f$, which is the case when energy conservation restricts scattering so that $|\mathbf{k}_f|\sim|\mathbf{k}_0|$ and the probe-system interaction is isotropic \cite{VanHove1954}. Anisotropic scattering, such as the dipole pattern in light scattering, can be taken into account by straightforward generalization.

The probability of scattering to any final wave-vector is highest when all the particles are localized to have zero relative position and the scattering probability becomes uniform. More generally the probability of scattering increases as the particles' relative positions become increasingly well localized, and observing a scattering event projects the state towards greater localization of the particles. 
The scattering distribution depends on $\left|\sum_{j=1}^N e^{i \mathbf{k}\cdot\mathbf{r}_j}\right|^2$, which is completely determined by the relative positions of pairs of particles. The information gained from a measurement of the scattering distribution is therefore limited to relative position and the measurement will then preserve superpositions of states that have the same set of relative position vectors.  Each state in such a superposition will give the same scattering pattern   and consecutive scattering measurements can push the system towards this type of superposition.
A similar effect is observed in arrays of Bose-Einstein condensates, where measurement of atoms coupled out of the condensates results in superposition of relative phase \cite{Dunningham2006a}. There, the relative phase localization was intrinsically limited by the number of atoms in the condensates, whereas here the number of scattering events and the resulting localization is unlimited.

Apart from scattering we must also consider the situation when the probe does not scatter from the system. Rather than leaving the system in the same state, the measurement of a non-scattering event also leads to a change in the state \cite{Rau2003a}. This is because non-scattering is more likely from certain particle distributions and detecting a non-scattering event projects the wavefunction of the system towards these configurations. For a position eigenstate $|\mathbf{R}\rangle$ the probability of non-scattering is determined by the conservation of probe particles, that is the probe particles must either be scattered or not scattered, giving 
\begin{equation}
P_{NS}(\mathbf{R}) = |A(\mathbf{R})|^2= 1 -  \frac{g^2}{4 \pi}\int d \Omega_{\mathbf{k}_f} \left|\sum_{j=1}^N e^{i \mathbf{k}\cdot\mathbf{r}_j}\right|^2,
\end{equation}
where the integration is over all scattering angles.
Detecting a non-scattering event then transforms the system state according to 
\begin{equation}
|\psi\rangle \longrightarrow \int d\mathbf{R} A(\mathbf{R})\phi(\mathbf{R}) |\mathbf{R}\rangle,
\label{eq:non_scatter_transform}
\end{equation}
and occurs with total probability $\int d\mathbf{R} |A(\mathbf{R})\phi(\mathbf{R})|^2$.
This projection favors basis states that have a lower probability of scattering and hence works in the opposite
direction to the projection following a scattering event. As we will see this prevents the system from always ending
in states with all the atoms localized with zero relative position.

Consecutive scattering and non-scattering events lead to a dynamic evolution of the many-body state. The dynamic scattering process can be simulated by the following procedure. Taking the initial state, we calculate the probability of non-scattering. A pseudo-random number is then generated to determine if non-scattering occurs. If it does then the projection in Eq.~(\ref{eq:non_scatter_transform}) is applied. If instead scattering occurs then the random number is used to determine the angle of scattering according to the probability density in Eq.~(\ref{eq:prob_scatter}), followed by projection of the state using Eq.~(\ref{eq:scatter_transform}). In either case the many-body state is then normalized and becomes the input state and the process repeats.

We now examine an example simulation that displays the key characteristics of the dynamic scattering process. We consider matter-wave scattering from a one-dimensional lattice of bosonic atoms in a sinusoidal potential \cite{Sanders2010a}, which can be achieved in an optical lattice \cite{Bloch2008a,*Lewenstein2007a}. We take the lattice to have $M$ sites and to be oriented along the $y$-axis, while the initial wave-vector of the probe is in the $x$-direction. For simplicity we only consider scattering within the $xy$-plane, where the full three dimensional scattering is a straightforward generalization.

At low temperature the atoms all reside in the lowest band of the lattice and the energy of the probing matter-waves can be arranged so that excitation of the atoms in the lattice to higher bands is negligible \cite{Sanders2010a}. The atomic field operator can then be reduced to an expansion in terms of lowest band Wannier functions $\hat{\Psi}(\mathbf{r}) = \sum_j w(\mathbf{r}-\mathbf{r}_j)$, where $r_j$ is the position of the $j$th lattice site \cite{Kittel}. We assume that the lattice potential is strong enough that the overlap between neighboring Wannier functions is negligible. The state of the system can be expressed in terms of a number basis $|\{n\}_u\rangle \equiv|\{n^{(u)}_j,j=1,\ldots,M\}\rangle$, where $n_j^{(u)}$ is the number of atoms at site $j$ and $u$ uniquely identifies each basis state. After the $m$th scattering event we expand the state as $|\Psi_m\rangle = \sum_u c_u^{(m)}|\{n\}_u\rangle$. Scattering from this state then occurs at angle $\theta$ to the $x$-axis with probability
\begin{equation}
P_m(\theta) = \frac{g^2}{2 \pi}\sum_u \left|I(\theta) c_u^{(m)}\sum_j e^{i \mathbf{r}_j \cdot \mathbf{k}(\theta)} n_j^{(u)}\right|^2.
\label{eq:ang_prob_density}
\end{equation}
where $I(\theta) = \int d\mathbf{r} e^{i \mathbf{k}(\theta) \cdot\mathbf{r}} |w(r)|^2$ and $\mathbf{k}(\theta)=k_0(1-\cos\theta,-\sin\theta)$.

Following a detection at $\theta$ the many-body state is projected into the new state
\begin{equation}
|\Psi_{m+1}\rangle = \frac{1}{\sqrt{\mathcal{N}}}\sum_u c_u^{(m)}\sum_j e^{i \mathbf{r}_j \cdot \mathbf{k}(\theta)} n_j^{(u)}|\{n\}_u\rangle,
\label{eq:theta_projection}
\end{equation}
where $\mathcal{N}$ normalizes the state.
We note that the number basis states are eigenstates of this projection, and the scattering process will preserve any state that begins in a basis state. Moreover, some of the basis states produce the same scattering pattern as the relative positions of pairs of atoms in the lattice is the same, for example in the $N=M=3$ case, $|2 0 1\rangle$ and $|1 0 2\rangle$ result in the same light scattering. Superposition of these states are partially preserved by the projection, in that the weight of each state in the superposition remains the same after scattering but the phase relationship is changed. For initial states that are the superposition of all the basis states, the projection gives higher weight to the basis states where the relative position is zero for each pair of particles, that is the state where $n_s = N$, for some lattice site $s$, and all other sites have zero occupancy. These states have the highest probability of scattering at any angle and detecting a scattering event makes it more probable that the atomic system is in one of these states. A sequence of scattering events with $\theta\neq 0$ then leads to a superposition of these states in the limit $m\rightarrow \infty$. 

%%%%%%%%%%%%%%%%%%%%%%% FIGURE %%%%%%%%%%%%%%%%%%%%%%%%%%%%%%%%%%%%%%%%%%%%%%%%%%%%%%%%%%%%%%%%%%%%%%%%%%%%%%%%%%%%%%%%%%%%%%
%%%%%%%%%%%%%%%%%%%%%%%%%%%%%%%%%%%%%%%%%%%%%%%%%%%%%%%%%%%%%%%%%%%%%%%%%%%%%%%%%%%%%%%%%%%%%%%%%%%%%%%%%%%%%%%%%%%%%%%%%%%%%

\begin{figure}
\centering
\includegraphics{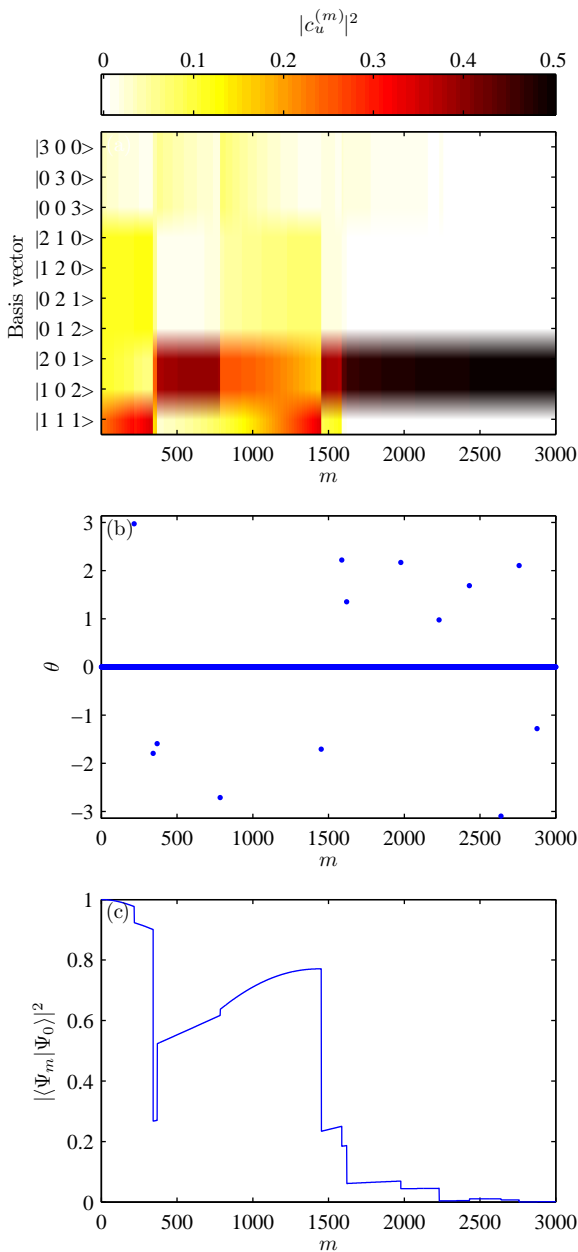}
\caption{Development of an atomic state caused by 3000 scattering events for a lattice with three sites and three atoms.  (a) Modulus squared of the basis coefficients of the state $|\Psi_m\rangle$ . (b) Detected events. (c) Overlap of the many-body state with the initial ground state $|\Psi_0\rangle$. Parameters used are $U/J = 0.05$, $g N=0.1$ and $k_0 = \pi/a$, where $a$ is lattice site separation and $U$ and $J$ are the parameters of the Bose-Hubbard model.}
\label{fig:develop_three_site}
%figure produced by Dynamic_light_scatter_with_non_scatter_sf_only_after_scat.m
\end{figure}

%%%%%%%%%%%%%%%%%%%%%%% FIGURE %%%%%%%%%%%%%%%%%%%%%%%%%%%%%%%%%%%%%%%%%%%%%%%%%%%%%%%%%%%%%%%%%%%%%%%%%%%%%%%%%%%%%%%%%%%%%%
%%%%%%%%%%%%%%%%%%%%%%%%%%%%%%%%%%%%%%%%%%%%%%%%%%%%%%%%%%%%%%%%%%%%%%%%%%%%%%%%%%%%%%%%%%%%%%%%%%%%%%%%%%%%%%%%%%%%%%%%%%%%%

In most cases, however, non-scattering events prevent the system from ending with all particles on one site, where these occur with probability
\begin{equation}
P_m^{NS} = \sum_u |c_u^{(m)} A_u|^2
\label{eq:prob_non_scat}
\end{equation}
and
\begin{equation}
|A_u|^2 = 1-\frac{g^2}{2\pi} \int_{-\pi}^\pi d\theta \left|I(\theta)\sum_j e^{i \mathbf{r}_j\cdot\mathbf{k}(\theta)} n_j^{(u)}\right|^2.
\end{equation}
Detecting a non-scattering event projects the state into the new state
\begin{equation}
|\Psi_{m+1}\rangle = \frac{1}{\sqrt{\mathcal{N}'}}\sum_u c_u^{(m)}A_u|\{n\}_u\rangle,
\label{eq:non_scat_projection}
\end{equation}
leading the state towards configurations that produce less uniform scattering distributions.

The evolution of the lattice many-body state can now be simulated using the above probabilities and projections.
To examine the dynamic process we look at the simple case of a three site lattice containing three atoms. In this case there are only ten basis states making it straightforward to track the development of the many-body state. We solve the Bose-Hubbard model numerically to find the initial many-body state of atoms in the lattice \cite{Jaksch1998b}. In Figure \ref{fig:develop_three_site} we show one realization of the dynamic scattering process for the three site lattice in the superfluid regime of the Bose-Hubbard model. For this example we have set the coupling constant $g N=0.1$ and as a result the vast majority of detection events result from non-scattering, however it only takes a few scattering events for the system to start favoring a particular final state. In this example the many-body state progresses toward a superposition of the states $|2 0 1\rangle$ and $|1 0 2\rangle$, two states which produce the same scattering pattern. As discussed above, continued scattering from this end superposition does not change the constituent basis vectors but does change the phase of the superposition. In Figure \ref{fig:develop_three_site}(c) we see that detection at non-zero angle quickly reduces the overlap of the many-body state with the original ground state. We see that the overlap makes quantum jumps when a scattering event occurs and gradually evolves due to non-scattering events.

By repeated simulation of the dynamic scattering process, we find that the simulations all settle into an end state after a number of scattering events. The end states are always superpositions of eigenstates of the scattering projection that produce the same scattering pattern. For the three site, three atom case there are four types of end state:
(1) superpositions of $|3 0 0\rangle$, $|0 3  0\rangle$, and $|0 0 3\rangle$, (2) superpositions of $|2 1 0\rangle$, $|1 2  0\rangle$, $|0 1 2\rangle$ and $|0 2 1\rangle$, (3) superpositions of $|2 0 1\rangle$, and $|1 0 2\rangle$, and (4) $|1 1 1\rangle$. Each superposition is associated with a different scattering distribution, which could be used to distinguish the superpositions in an experiment.

\begin{figure}
\centering
\includegraphics{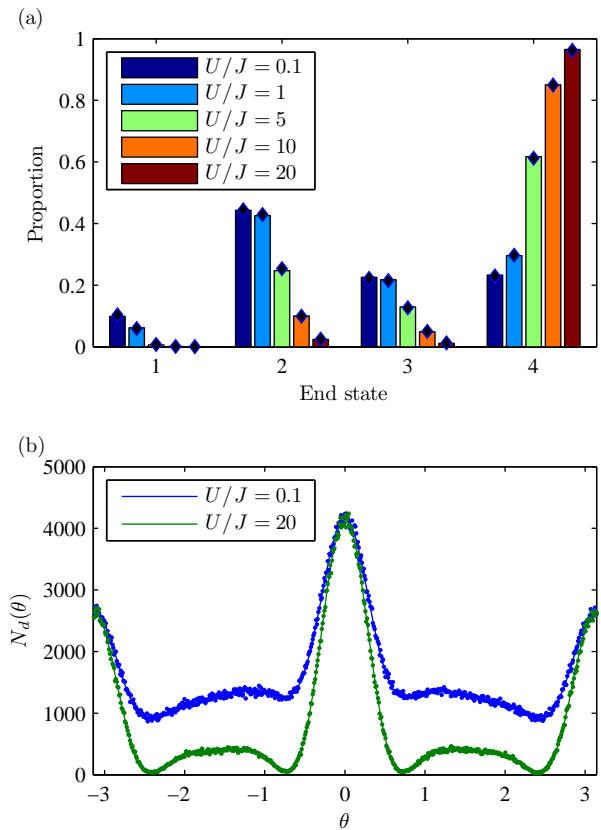}
\caption{(a) Proportion of 10000 simulations that end in a final state superposition of (1) $|3 0 0\rangle$, $|0 3  0\rangle$, and $|0 0 3\rangle$, (2) $|2 1 0\rangle$, $|1 2  0\rangle$, $|0 1 2\rangle$ and $|0 2 1\rangle$, (3) $|2 0 1\rangle$, and $|1 0 2\rangle$, and (4) $|1 1 1\rangle$ at various values of $U/J$. The initial probabilities of finding these basis states in the initial ground state as shown by the diamonds. (b) (points) Number of event detections $N_d(\theta)$ after 10000 simulations of 1000 detection events each, where the angular range is divided into 600 bins for event counting. (lines) Scattering distributions predicted from the initial ground states.  Parameters used are $gN=0.5 $ and $k_0 = \pi/a$.}
\label{fig:state_photon_dist}
%figure produced by Plot_multiple_sim_end_state_distributions_N.m
\end{figure}

In the Mott insulator regime of the Bose-Hubbard model the initial ground state is dominated by the $|1 1 1\rangle$ basis state and the most likely outcome of the dynamic process is to drive the state completely into the $|1 1 1\rangle$ state. 
In fact we find that the probability the end state of the dynamical scattering process is $|1 1 1\rangle$ is equal to $|\langle 1 1 1|\Psi_0\rangle|^2$, where $|\Psi_0\rangle$ is the initial state. This generalizes for final states that are superpositions, for example the end state superposition of basis states $|2 0 1\rangle$ and $|1 0 2\rangle$ occurs with probability $|\langle 2 0 1|\Psi_0\rangle|^2 + |\langle 1 0 2|\Psi_0\rangle|^2$.
In Figure \ref{fig:state_photon_dist}(a) we plot the proportion with which each possible final state superposition is represented in an ensemble of 10000 simulations for various ratios of the on-site interaction strength $U$ and tunneling energy $J$.
We see that the proportions match the initial probabilities of finding the relevant basis states in the initial ground state. The Mott state almost always ends in the $|1 1 1\rangle$ state, while for the superfluid, the end states are distributed across all possible outcomes. These results are independent of the coupling parameter $g$, which only affects the rate at which the state approaches an end state.

%%%%%%%%%%%%%%%%%%%%%%%%%%%%FIGURE%%%%%%%%%%%%%%%%%%%%%%%%%%%%%%%%%%%%%%%%%%%%%%%%%%%%%%%%%%%%%%%%%%%%%%%%%%%%%%%%%%%%%%%%%%%%%%%%%%%%%%%%%%%%%%%%%%%%%%%%%%%%%%%%%%%%%%%%%%%%%%%%%%%%%%%%%%%%%%%%%%%%%%%%%%%%%%%%%%%%%%%%%%%%%%%%%%%%%%%%%%%%%%%%%%%%%%%%%%%%%%

%%%%%%%%%%%%%%%%%%%%%%%%%%%%FIGURE%%%%%%%%%%%%%%%%%%%%%%%%%%%%%%%%%%%%%%%%%%%%%%%%%%%%%%%%%%%%%%%%%%%%%%%%%%%%%%%%%%%%%%%%%%%%%%%%%%%%%%%%%%%%%%%%%%%%%%%%%%%%%%%%%%%%%%%%%%%%%%%%%%%%%%%%%%%%%%%%%%%%%%%%%%%%%%%%%%%%%%%%%%%%%%%%%%%%%%%%%%%%%%%%%%%%%%%%%%%%%%

Because the final state proportions are the same as the initial basis state probabilities, we expect that the scattering intensity distribution will equal that given by the initial ground state scattering probability distribution \emph{if} we average over an ensemble of scattering experiments. This occurs even though the initial state can be completely changed in the scattering process. 
In Figure \ref{fig:state_photon_dist}(b) we show the intensity distributions for two systems, one in the superfluid regime and the other in the Mott regime, for an ensemble of 10000 scattering simulations with 1000 detection events each. The scattering intensity distributions for each ensemble match the scattering probability for the initial ground states, despite that in the superfluid case the individual realizations do not give the correct distribution. For light scattering from atoms in a lattice we expect a similar result, however in this case contributions from higher bands must be accounted for \cite{Douglas2011a}. Light scattering from optical lattices into a cavity has also been suggested as a method for creating particular many-body states \cite{Mekhov2009a,*Mekhov2009b}.

In this treatment we have not included the effect of time development due to the atomic Hamiltonian, under the assumption that the scattering process occurs on a time scale short compared to the tunneling and interaction time scales. For slower scattering this could be taken into account by evolving the system using the atomic Hamiltonian in between the scattering events.

Our example of scattering from atoms in a lattice shows how scattering and measurement pushes the atomic many-body state towards eigenstates of the scattering projection. In some cases this leads to superpositions of many-body states in position space, where the constituent states give the same scattering pattern. These states are robust under continued scattering, unlike other more fragile superpositions where interactions with the environment lead to decoherence \cite{Zurek1991a}. The constituents of the superposition are revealed by the scattering pattern that results after multiple scattering events, and scattering could be used to generate these entangled states in a non-deterministic fashion. It is also apparent that scattering distributions reveal the initial state scattering distribution, even when the initial state is significantly perturbed, provided the distributions are averaged over a number of experimental runs.

%%%%%%%%%%%%%%%%%%%%%%%%%%%%%%%%%%%%%%%%%%%%%%%%%%%%%%%%%%%%%%%%%%%%%%%%%%%%%%%%%%%%%%%%%%%%%%%%%%%%%%%%%%%%%%%%%%%%%%%%%%%%%%%%%%%%%%%%%%%%%%%%%%%%%%%%%%%%%%%%%%%%%%%%%%%%%%%%%%%%%%%%%%%%%%%%%%%%%%%%%%%%%%%%%%%%%%%%%%%%%%%%%%%%%%%%%%%%%%%%%%%%%%%%%%%%%%%%

%

\end{document}